\documentclass[twocolumn, prl,superscriptaddress,amsfonts,showpacs,floatfix]{revtex4} 

\usepackage{graphicx}
\usepackage{dcolumn}
\usepackage{bm}
\usepackage{color}

\begin{document}

\preprint{APS/123-QED}

\title{Field-Orientation Dependence of  Low-Energy Quasiparticle Excitations\\
 in the Heavy-Electron Superconductor UBe$_{13}$ 
}



\author{Yusei Shimizu}
\email{yuseishimizu@issp.u-tokyo.ac.jp}
\affiliation{Institute for Solid State Physics (ISSP), University of Tokyo, Kashiwa, Chiba 277-8581, Japan}
\author{Shunichiro Kittaka}
\affiliation{Institute for Solid State Physics (ISSP), University of Tokyo, Kashiwa, Chiba 277-8581, Japan}
\author{Toshiro Sakakibara} 
\affiliation{Institute for Solid State Physics (ISSP), University of Tokyo, Kashiwa, Chiba 277-8581, Japan}
\author{ Yoshinori Haga}
\affiliation{Advanced Science Research Center, Japan Atomic Energy Agency, Tokai, Ibaraki,  319-1195,  Japan. }
\author{ Etsuji Yamamoto}  
\affiliation{Advanced Science Research Center, Japan Atomic Energy Agency, Tokai, Ibaraki,  319-1195,  Japan. }
\author{  Hiroshi Amitsuka} 
\affiliation{Graduate School of Science, Hokkaido University, Sapporo, Hokkaido,  060-0810, Japan.}
\author{ Yasumasa Tsutsumi} 
\affiliation{Condensed Matter Theory Laboratory, RIKEN, Wako, Saitama, 351-0198, Japan. }
\author{ Kazushige Machida }     
\affiliation{Department of Physics, Okayama University, Okayama,  700-8530, Japan. }


\date{\today}%

\begin{abstract}
Low-energy quasiparticle excitations in the superconducting (SC) state of UBe$_{13}$
 were studied by means of  specific-heat ($C$) measurements in a rotating field.
Quite unexpectedly, the magnetic-field dependence of $C(H)$ is linear in $H$  
 with no angular dependence  
at low fields 
  in the SC state,  implying 
  that the gap is fully open over the Fermi surfaces, in stark contrast to the previous expectation.
In addition, a characteristic cubic anisotropy of $C(H)$ was observed above 2~T with a maximum (minimum) for $H$ $||$ $[001]$ ($[111]$) 
  within the $(1\bar{1}0)$  plane,  both in the normal as well as in the SC states.
This oscillation possibly originates from the anisotropic response of the heavy quasiparticle bands, and might be a key to understand the unusual properties of UBe$_{13}$.
\end{abstract}

\maketitle


Three decades have  passed since the discoveries of  superconductivity in
CeCu$_{2}$Si$_{2}$ \cite{Steglich_PRL_1979} and UBe$_{13}$ \cite{Ott_PRL_1983}.
It is widely accepted that their pairing mechanisms are unconventional, fundamentally because the 
effective Fermi temperatures of these systems ($\sim$10~K) are, like other heavy-electron  superconductors, much lower than the Debye temperatures~\cite{Steglich_PRL_1979}.
Determination of the gap symmetry of the heavy-electron superconductors is, however, by no means an easy task,
and there are only a few heavy-electron superconductors whose superconducting (SC) gap structures are fully elucidated.
For instance, the SC pairing symmetry of CeCu$_{2}$Si$_{2}$ has not been clarified until very recently~\cite{Kittaka_PRL_2014}.

\color{black}
In this Letter,
 we focus on  a  cubic heavy-electron superconductor UBe$_{13}$.
 Despite extensive studies over 30 years,  the  nature of superconductivity in UBe$_{13}$ is still elusive.
The $^9$Be-NMR-Knight shift has been reported to be invariant below the superconducting transition temperature $T_{\mathrm{c}}\approx 0.86$~K \cite{Tien_PRB_1989,Tou07},
 suggesting an odd-parity pairing.
However,  
   $\mu^{+}$SR-Knight-shift experiment indicates a significant decrease of the static susceptibility below $T_{\mathrm{c} }$
 \cite{Sonier_PhysicaB_2003}, conflicting with the NMR results. 
Regarding the gap structure,
 whereas the specific-heat $C(T)$ \cite{Ott_PRL_1984} and the magnetic penetration depth \cite{Einzel_PRL_1986} experiments suggest the presence of point nodes,
 the  NMR spin-relaxation rate  \cite{MacLaughlin_PRL_1984} and the ultrasound attenuation \cite{Golding_PRL_1985}  
 are rather indicative of line nodes.  
Although there is compelling evidence for unconventional pairing, the SC gap symmetry in UBe$_{13}$ thus remains  undetermined.

Another important issue with respect to the SC state in UBe$_{13}$ is a feature observed in thermodynamic quantities such as $C(H)$~\cite{Ellman_PRBR_1991, Kromer_PRL_1998, Walti_PRBR_2001},  dc magnetization $M(H)$~\cite{YShimizu_PRL_2012} as well as the thermal expansion~\cite{Ellman_PRBR_1991}
  at  fields below $\sim$4~T,  constituting a line of anomaly (``$B^*$ anomaly'') in the $H-T$ phase diagram.
Whereas the origin of this anomaly is still unresolved yet, it has been discussed as a precursor~\cite{Kromer_PRL_1998} of the second phase transition below $T_{\rm c}$ observed in 
U$_{1-x}$Th$_{x}$Be$_{13}$ (0.019 $<$ $x$ $<$ 0.045)~\cite{Ott_PRBR_1985}.

Normal state of UBe$_{13}$ is also highly unusual.
 It exhibits non-Fermi-liquid (NFL) behavior down to very close to $T_{\rm c}$ as revealed by
 electrical resistivity, specific heat \cite{Mayer_PRB_1986, Gegenwart_PhysicaC_2004},  and magnetic susceptibility \cite{McElfresh_PRB_1993}.
The origin of NFL behaviors in UBe$_{13}$ remains unclear, 
 and several possibilities have been discussed so far.
These include
 quadrupolar Kondo effect  with $\Gamma_{3}$-crystalline-electric-field  ground state for $5f^{2}$ (U$^{4+}$, $J$=4) configuration \cite{Cox_PRL_1987},
 an antiferromagnetic quantum-critical point  induced by a magnetic field\cite{Gegenwart_PhysicaC_2004, Schmiedeshoff_PhilosMagazine_2009},
and  a competition between Kondo-Yosida   and   $\Gamma_{1}$-crystalline-electric-field singlets for $5f^{2}$ 
 configuration~ \cite{Nishiyama_JPSJ_2011}.
 Since the SC state apparently emerges out of the NFL state, its understanding is crucial in elucidating the pairing mechanism in UBe$_{13}$.

\begin{figure}[!htb]
\begin{center}
\includegraphics[width=8.5cm]{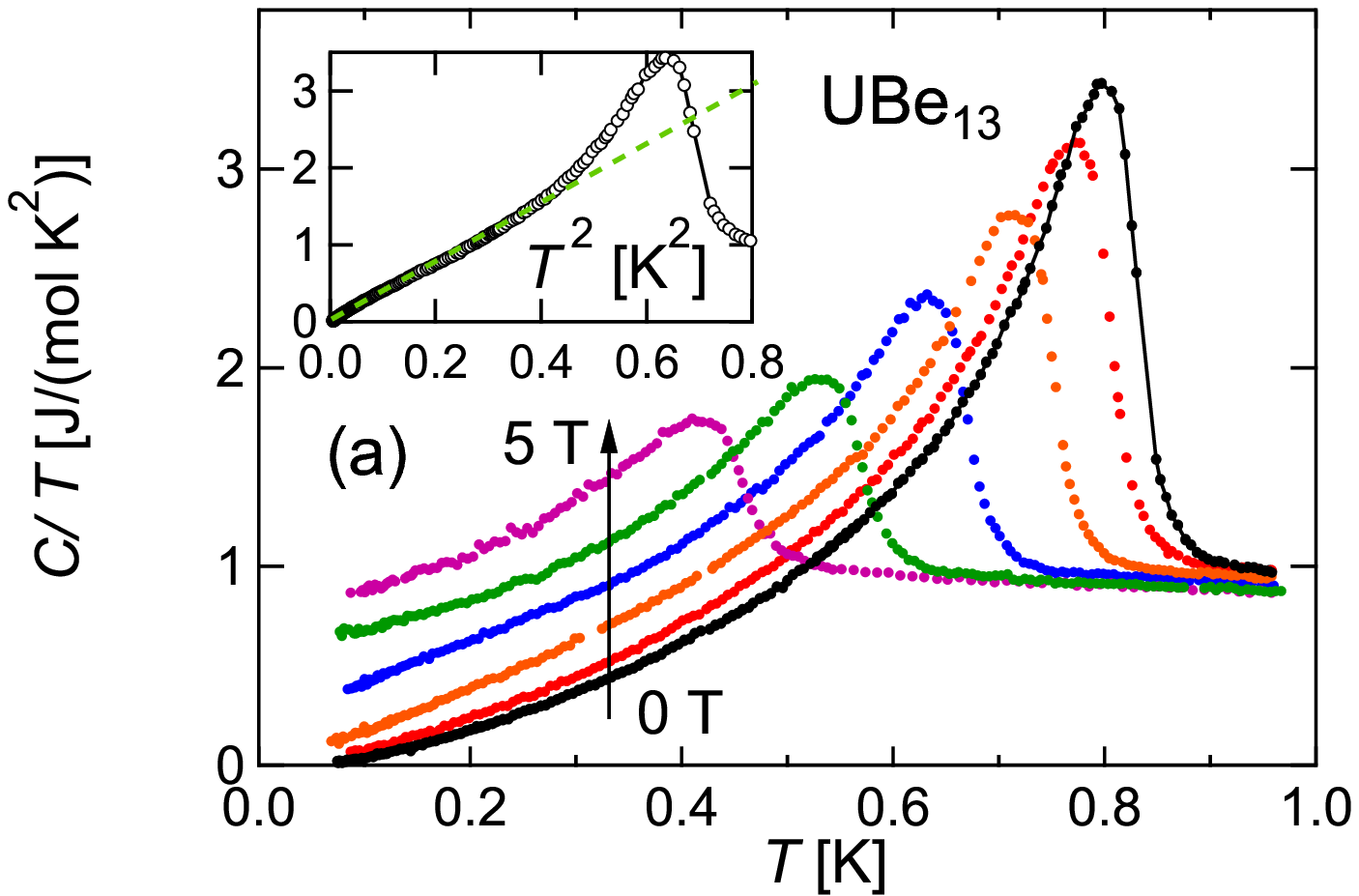}
\includegraphics[width=8.5cm]{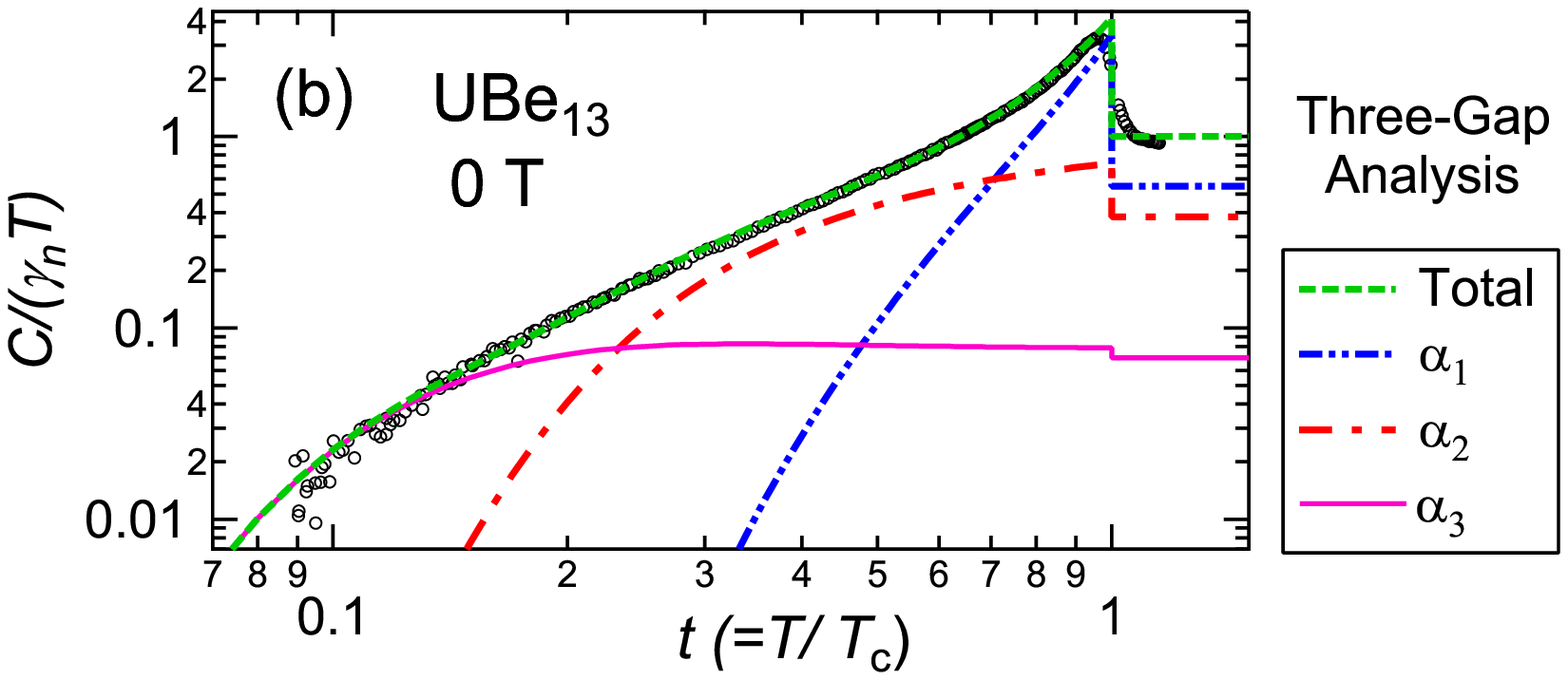}
\includegraphics[width=4.2cm]{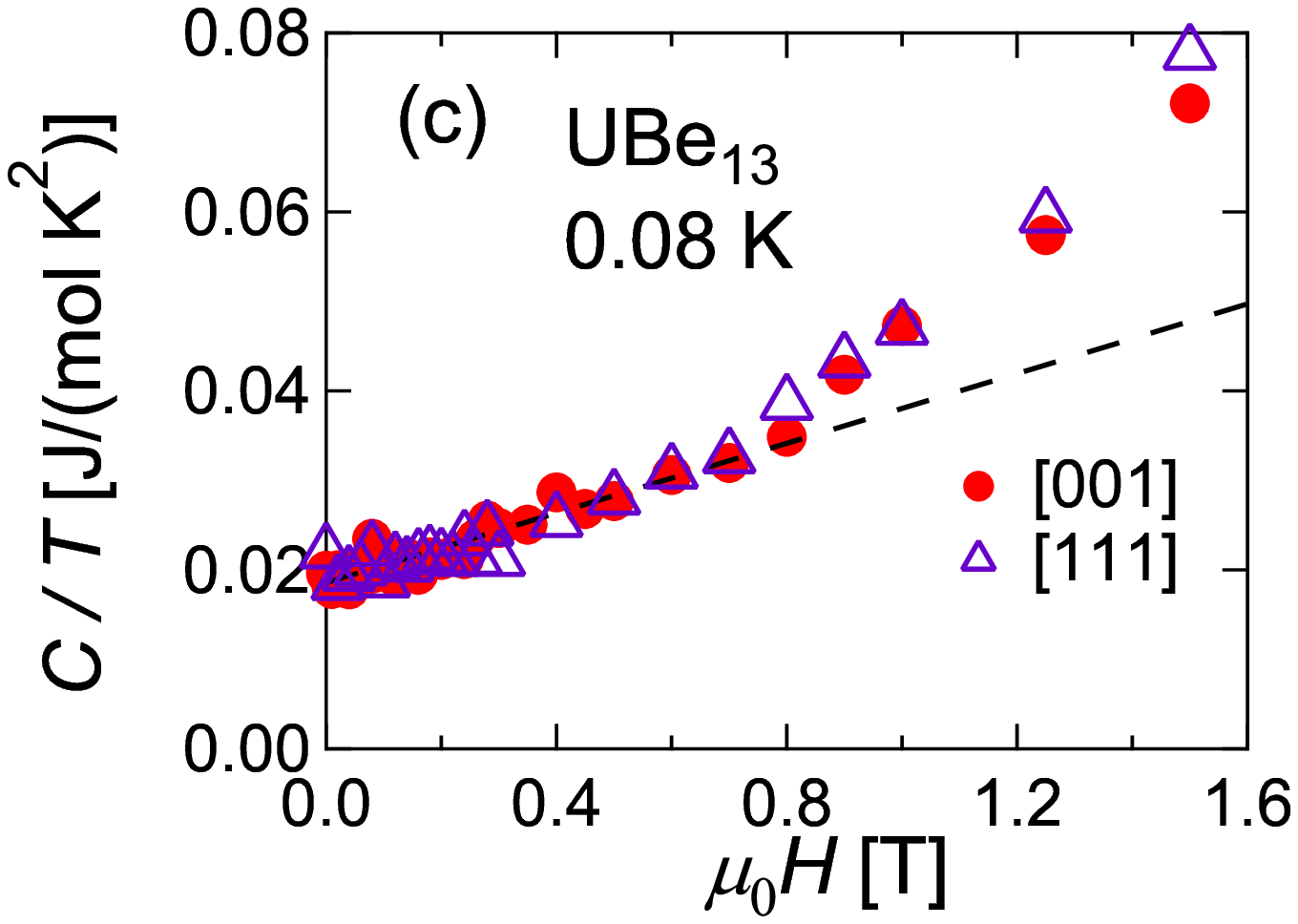}
\includegraphics[width=4.2cm]{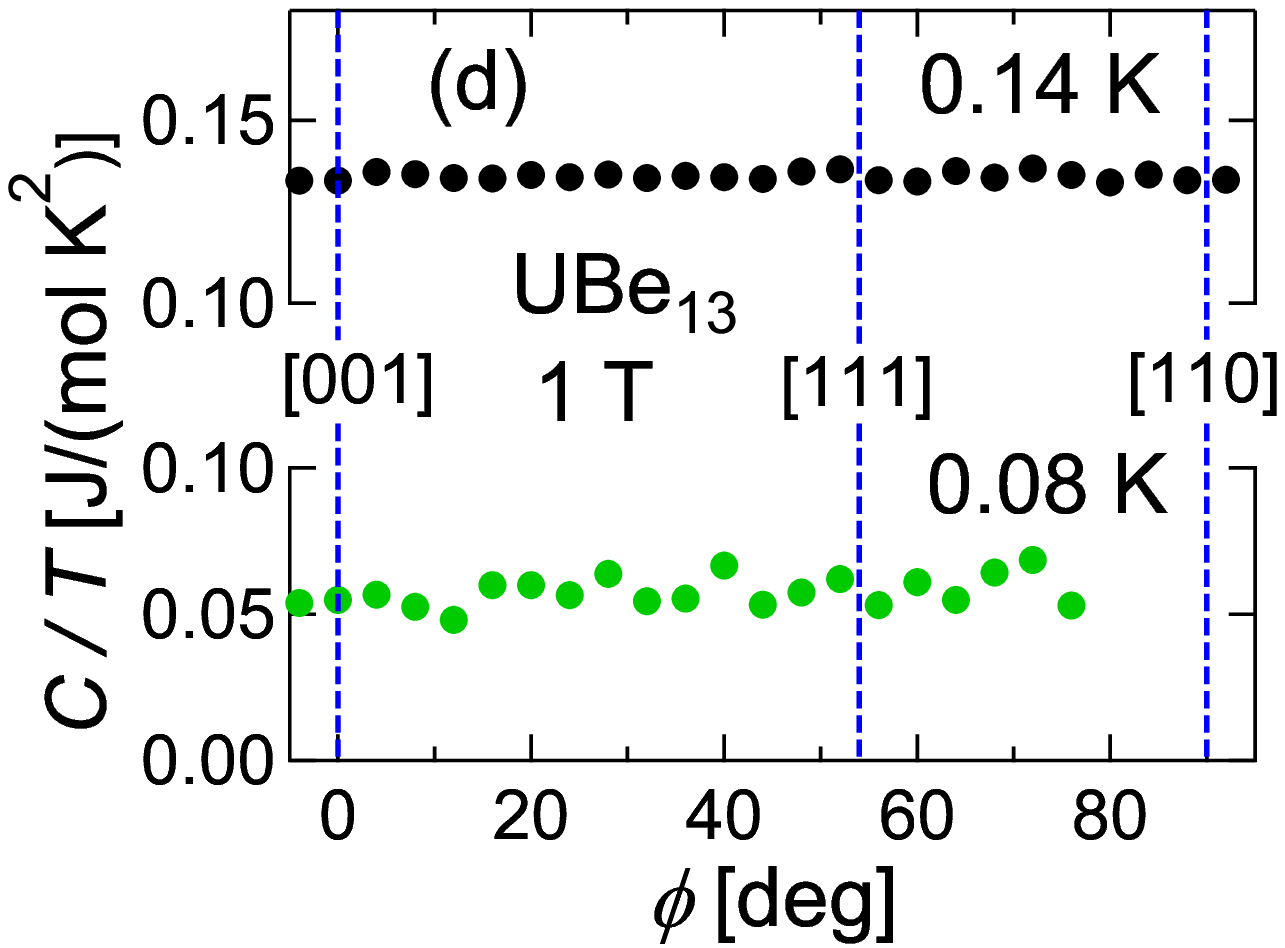}
\caption{ (Color online) (a) $C(T)/T$ of UBe$_{13}$ at $\mu_0H=0$, 1, 2, 3, 4, and 5~T for $H || [001]$. 
The inset shows $C(T)/T$ vs $T^{2}$ plot for $H=0$. 
 (b) The result of three-band full-gap analysis for $C(T)/T$ ($H=0$) in log-log scale. 
The parameters are: $\alpha_{1} = 1.9$, $\alpha_{2} = 0.8$, $\alpha_{3} = 0.3$, and 
 $\gamma_{1}$ : $\gamma_{2}$ : $\gamma_{3} $ = 55 : 38 : 7.
(c)  $C(H)/T$ at $T=0.08$~K for $H ||[001]$ (solid circles) and $H ||[111]$ (open triangles) as a function of $H$ in the low-field region.
The dashed line is a linear fit to the data below $\sim$0.8~T for $H ||[001]$.
(d)  $C(\phi)/T$ in a field of 1~T rotated in the  $(1\bar{1}0)$  plane, measured at 0.08, and 0.14~K. 
 }
\end{center}
\end{figure}

In order to gain more insight into 
 the SC gap symmetry  as well as the normal state,
 we performed  specific-heat measurements of UBe$_{13}$ at low temperatures   
 down to $\sim$75~mK in magnetic fields up to 5~T.
 The single crystal of UBe$_{13}$, used in the present study  was prepared by an Al-flux method
 \cite{Haga_Physica_1999}. This is the same crystal as used in the previous dc magnetization study
 \cite{YShimizu_PRL_2012}. 
The specific heat $C$ was measured  by a standard quasi-adiabatic heat-pulse method.
Field-angular  dependences $C(H, \phi)$ were  measured with $H$ rotating in the  $(1\bar{1}0)$ crystal plane
 that includes  three principal directions $[001]$, $[111]$, and $[110]$.
The angle $\phi$ is measured from the $[001]$ axis.

Figure 1(a) shows $C(T)/T$ curves measured in various magnetic fields up to 5 T.
The zero-field data are also plotted in Fig.~1(b) in log-log scale.
There is no Schottky contribution from $^9$Be nuclei, owing to their long nuclear spin-relaxation time of the order of 10$^3$~sec  
 \cite{MacLaughlin_PRL_1984},  much longer than our measuring time (10$^2$~sec) of the specific heat.
The inset of Fig. 1(a) shows the $C(T)/T$ vs $T^2$ plot; 
 $C(T)$ below $\sim$ 0.6~K behaves like  $T^{3}$ as previously reported  
 \cite{Ott_PRL_1984}.
Note that the residual density of states, $C(T)/T|_{T \rightarrow 0}$, is very small.

Magnetic-field dependence of the specific heat and its anisotropy 
 in low fields reflect quasiparticle excitations  within the SC gap 
 \cite{Volovik_JETP_Lett_1993, Vekhter_PRBR_1999, Sakakibara_JPSJ_2007}.
In the case of line nodes, 
  $C(H) \propto (H/H_{\mathrm{c2} })^{1/2}$ is expected 
 \cite{Volovik_JETP_Lett_1993,
 Vekhter_PRBR_1999, Sakakibara_JPSJ_2007}, 
  whereas for point nodes, $C(H) \propto \frac{H}{H_{\mathrm{c2} } } \mathrm{ln} \frac{H}{H_{\mathrm{c2} } } $ 
 \cite{Volovik_JPhysC_21_1988}, 
 or $C(H) \propto (H/H_{\mathrm{c2} })^{0.64}$ 
 \cite{Miranovic_PRB_2003}.
In either case,  
 the 
 field dependence of $C/T$ should exhibit  
a convex upward curvature at low fields.
For a clean isotropic $s$-wave superconductor, on the other hand,
$C(H)/T \propto H$  at low fields because  low-energy quasiparticles are mainly confined  
in  vortices whose density increases in proportion to $H$ 
 \cite{Nakai_PRBR_2004}.
Figure 1(c) shows  $C(H)/T$  of UBe$_{13}$  below 1.6~T
 for $H || [001]$ and  $H || [111]$ measured  at 0.08~K.
Surprisingly, the low-field $C(H)/T$ curve is rather \textit{linear} in $H$, suggesting the absence of nodal quasiparticles.
Note that there is \textit{no} anisotropy in this $H$-linear behavior of $C(H)$ below $\sim$1~T
  between [001] and [111] directions within an experimental accuracy
 [Fig.~1~(c)].
The absence of the anisotropy is further confirmed by $C(\phi)/T$ obtained  in a field of 1~T rotated in the $(1\bar{1}0)$ crystal plane at $T=0.08$ and 0.14~K [Fig. 1(d)];
there is \textit{no} significant angular variation in $C(\phi)/T$, implying that the $C(H)/T \propto H$  behavior holds for all directions.

\begin{centering}
\begin{figure}
\begin{minipage}{5.1cm}
\includegraphics[width=5.1cm]{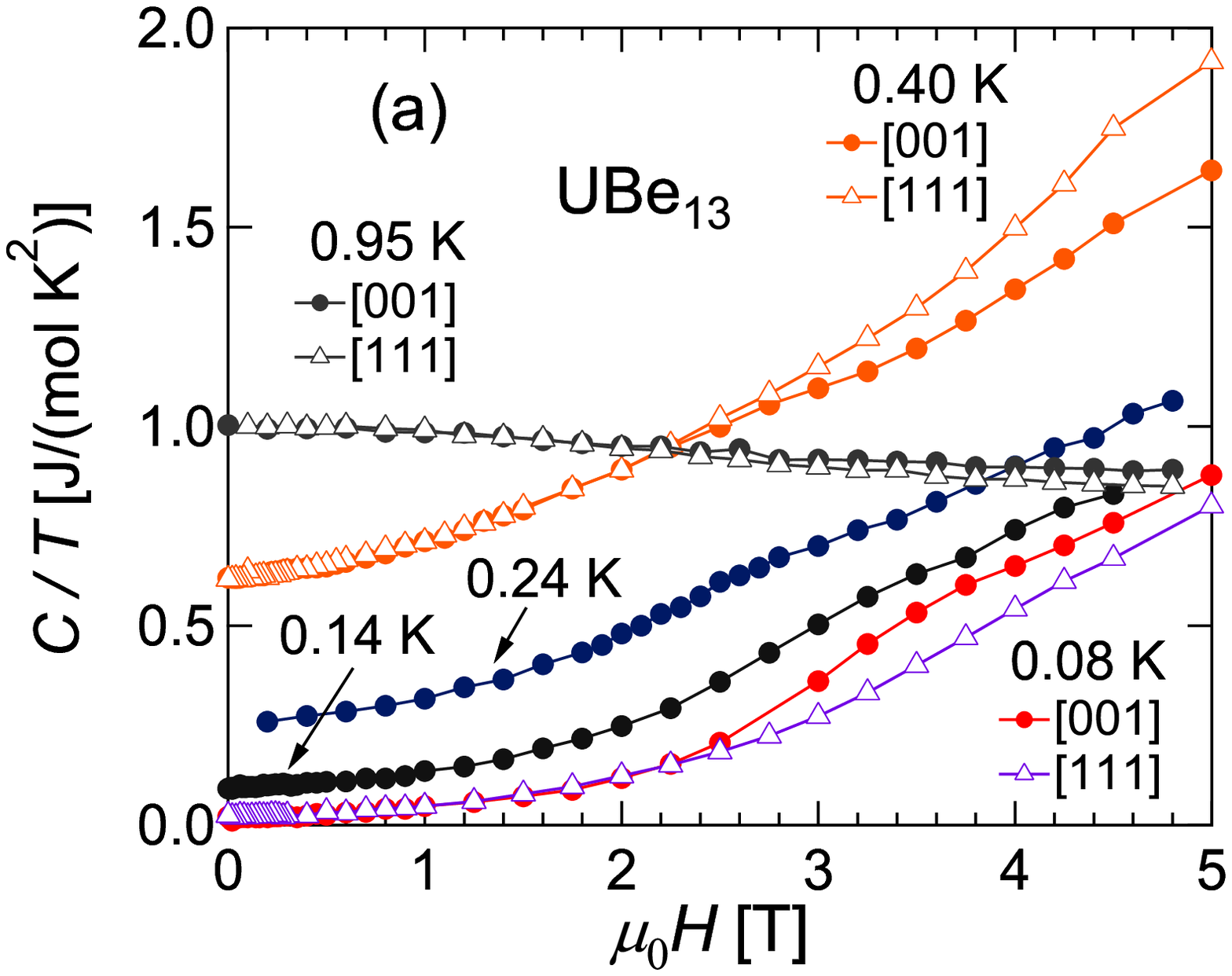}
\end{minipage}
\begin{minipage}{3.4cm}
\includegraphics[width=3.4cm]{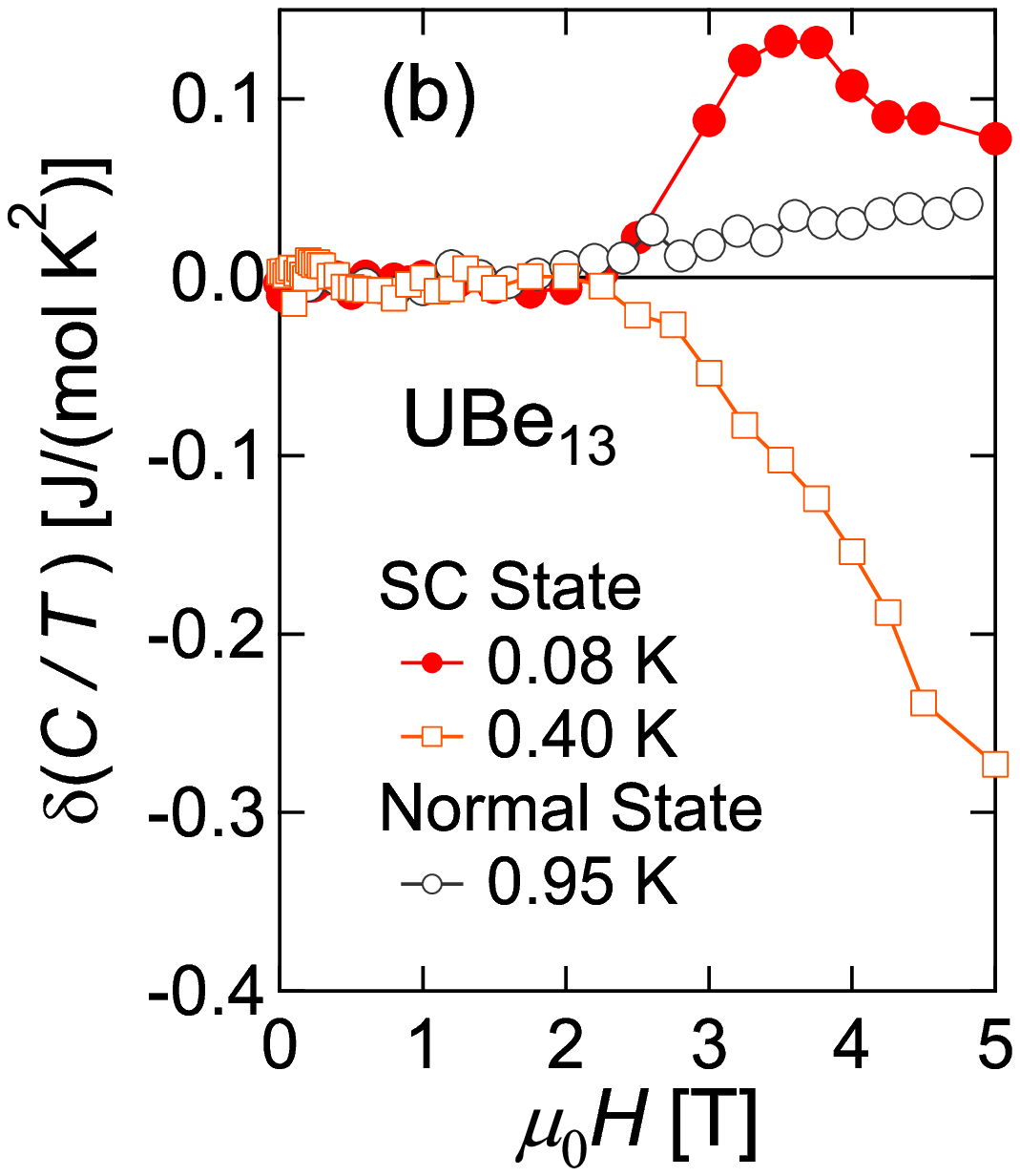}
\end{minipage} 
\caption{ (Color online) (a) Magnetic-field dependence of  $C(H)/T$ up to 5~T
 for $H$$||$$[001]$ (solid circles) and $H$$||$$[111]$ (open triangles) measured at $T=0.08$, 0.14, 0.24,  0.40, and 0.95~K.
 (b) $\delta (C/T)$ $\equiv$ $(C_{[001]}$$-$$C_{[111]})/T$ as a function of $H$, obtained
  at $T=0.08$ (solid circles), 0.40 (open squares), and 0.95~K (open circles).
  }
\end{figure}
\end{centering}

We would like to emphasize that the linear slope of $C(H)/T$ in Fig.~1(c) is unusually small. 
In this regard, it has been argued that the line-nodal or point-nodal sublinear dependences in $C(H)/T$ described above would be smeared out at a finite temperature $T/T_{\rm c}\!\!>\!\!\sqrt{H/H_{\rm c2}}$~\cite{Volovik97}.
Even in such case, however, the rate of increase of $C(H)/T$ should be greater than that governed by
the localized-quasiparticle contribution from  vortex cores approximated as  $C(H)/T$=$(C_{\mathrm{n}}/T) (H/H_{\mathrm{c2} }^{\mathrm{orb} }(0) )$, where $\mu_0H_{\mathrm{c2} }^{\mathrm{orb}}$ (=25~T~\cite{Maple_PRL_1985}) denotes
  the orbital-limiting field.
Assuming the normal-state value $C_{\mathrm{n}}/T$ at $T$$\sim$0 to be 1.1~J$\cdot$mol$^{-1}\cdot$K$^{-2}$ by taking into account an entropy balance, 
we estimate the slope $(C_{\mathrm{n}}/T) /(\mu_0H_{\mathrm{c2} }^{\mathrm{orb} }(0) )$ of the ordinary vortex core contribution  to be 
0.044~J$\cdot$mol$^{-1}\cdot$K$^{-2}$T$^{-1}$.
The observed initial slope 
 in Fig. 1(c) is 0.02~J$\cdot$mol$^{-1}\cdot$K$^{-2}$T$^{-1}$, a factor of two smaller than this; apparently, there is a significant deficiency of  quasiparticles.  
 We will come back to this point later.

Figure 2(a) shows the field variation of $C(H)/T$ up to 5~T  for $H$$||$$[001]$ measured at $T=0.08$, 0.14, 0.24,  0.40, and 0.95~K (closed symbols).
We also plot the data for $H$$||$$[111]$ measured at $T=0.08$, 0.40, and 0.95~K (open symbols).
$C(H)/T$ curve for $H$$||$$[001]$  in the SC state at $T= 0.08$~K exhibits a strong upturn above $\sim 2$~T. This behavior is quite reminiscent of a superconductor with a strong Pauli paramagnetic effect, as observed for CeCu$_2$Si$_2$~\cite{Kittaka_PRL_2014}.
 Note that a weak hump appears in $C(H)/T$ above $\sim$3~T,
 whose position moves to lower fields with increasing $T$.
This hump has been known as the 
``$B^{*}$ anomaly'' \cite{Ellman_PRBR_1991, Kromer_PRL_1998, Walti_PRBR_2001, YShimizu_PRL_2012}.
We observe that this anomaly in $C(H)$ is clearer for  $H$$||$$[001]$ than  for $H$$||$$[111]$.
Accordingly, a substantial anisotropy develops in  $C(H)$ above this field.
An anisotropy has also been observed by  dc magnetization curves above $B^{*}$   
 \cite{YShimizu_DrThesis_2012, YShimizu_JPSconfProc_2014}.
In order to display the evolution of the anisotropy in $C(H)/T$,  we plot  in Fig.~2(b)
the difference $\delta (C/T)$ $\equiv$ $(C_{[001]}$ $-$ $C_{[111]})/T$ at $T=0.08$, 0.40, and 0.95~K,  
where  $C_{[001]}$ ($C_{[111]}$) denotes the specific heat for $H$$||$$[001]$ ($H$$||$$[111]$).
For $T= 0.08$~K,  $\delta (C/T)$ shows a distinct positive peak 
 around 3.5~T due to the $B^{*}$ anomaly. 
 At 0.40~K, $\delta (C/T)$ changes the sign and  shows a monotonic decrease with increasing field above 2~T,
  reflecting the anisotropy of $H_{\mathrm{c2} }$.
 In the normal state at 0.95~K,
 $\delta (C/T)$ turns positive again and increases monotonically with increasing field above $\sim$2~T.
It is also remarkable that the normal-state $C/T$ is substantially suppressed in a field of 5~T for both directions.

\begin{centering}
\begin{figure}[t]
\includegraphics[width=4.2cm]{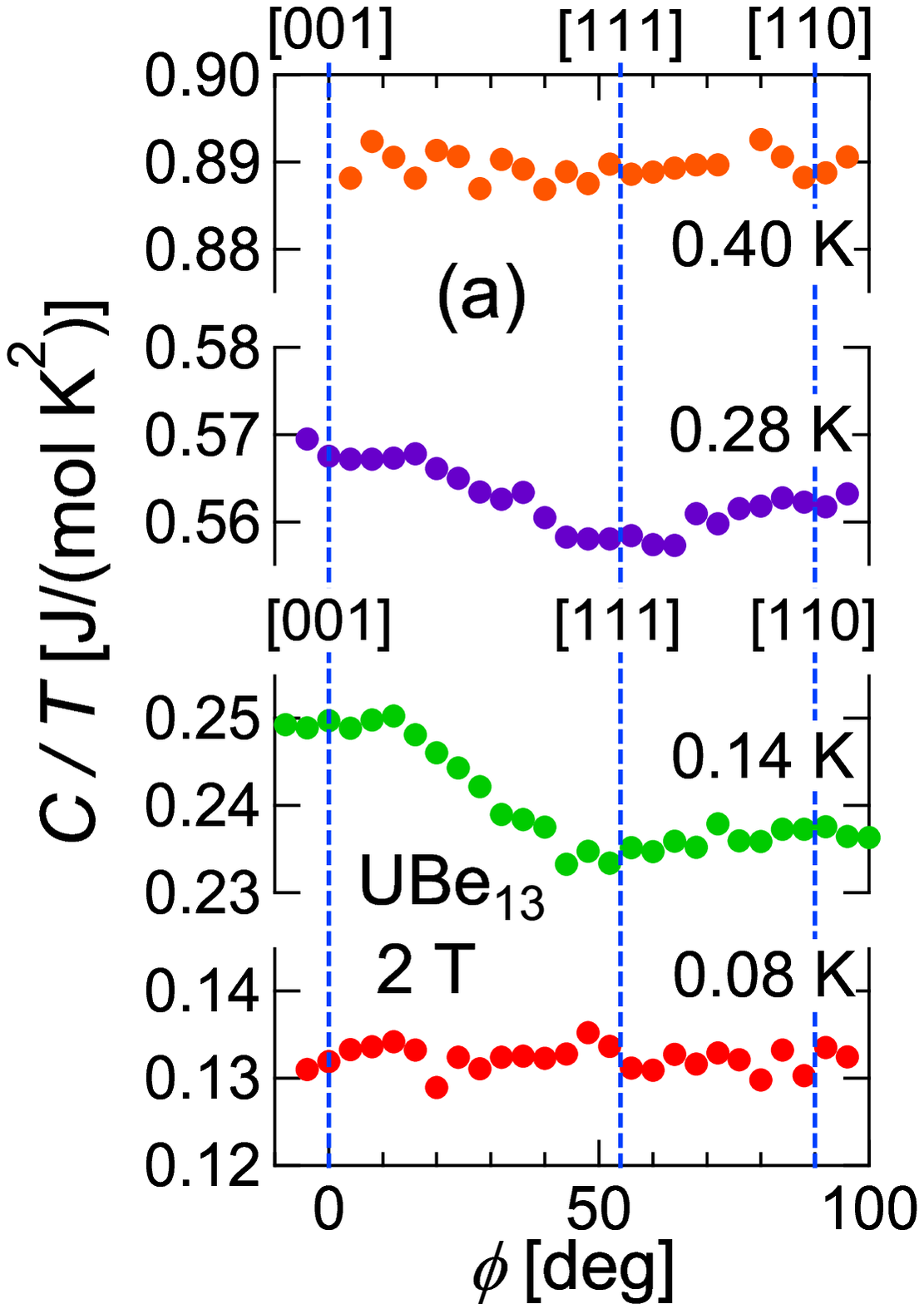}
\includegraphics[width=4.2cm]{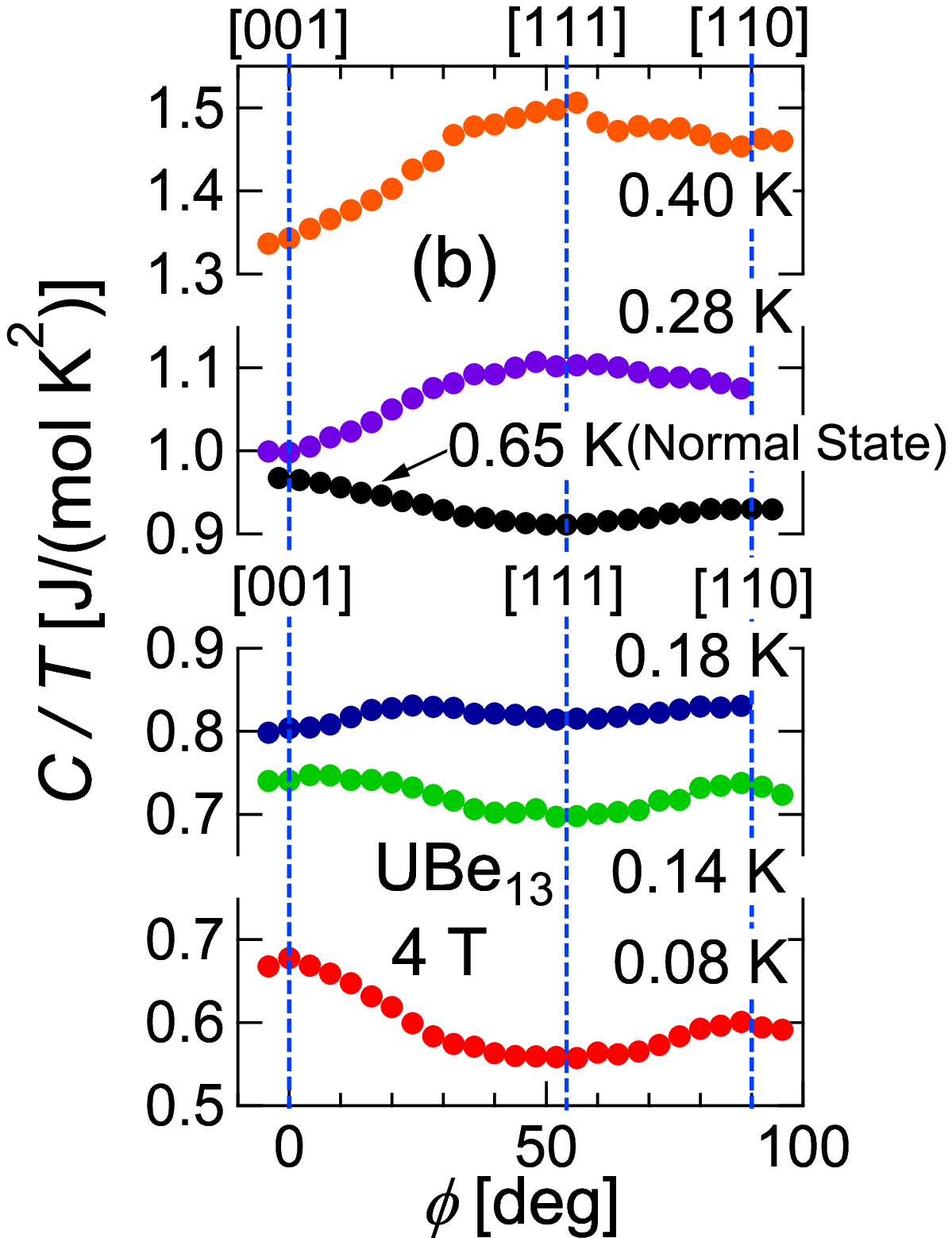}
\caption{   Angular dependences of $C(H, \phi)/T$, measured in (a) $\mu_0H=2$~T, and (b) 4~T 
  at $T=0.08$, 0.14, 0.28, and 0.40~K.
Data for $T=0.65$~K and $\mu_0H=4$~T in the normal state are also plotted.
 }
\end{figure}
\end{centering}


In Fig.~3, we  show the field-angle dependences of $C/T$ in the $(1\bar{1}0)$  plane measured in a magnetic field of  (a) 2~T, and (b) 4~T.
For $\mu_0H=2$~T, an appreciable angular variation can be seen at $T=0.28$, and 0.14 K,
  with a maximum (minimum) at $[001]$ ($[111]$) and a local maximum at $[110]$, i.e. $C_{[111] }$ $<$ $C_{ [110] }$ $<$ $C_{ [001] } $. 
Hereafter, we call this type of the angular variation as ``type-I oscillation''.
At this field, the oscillation disappears for $T=0.08$~K reflecting the absence of nodal quasiparticles.
When the magnetic field is increased to 4~T, the type-I oscillation appears even at a low temperature of $T=0.08$~K,
with a huge relative amplitude of nearly 25\%.
Such a large anisotropy cannot be ascribed to nodal quasiparticles.
With increasing $T$, the sign of the oscillation changes for $T=0.28$ and 0.40~K.
We call this reversed angular variation as ``type-II oscillation''.
The reversed oscillation at these higher temperatures is probably due to an anisotropy of $H_{\mathrm{c2}}$.
Note that 
 the  amplitude of  both the type-I and type-II oscillations at 4 T 
 is nearly 10-times larger than that at 2 T; the anisotropy strongly develops with increasing $H$ as can be seen from Fig.~2(b).
For $T= 0.14$ and 0.18~K,  $C(H, \phi)/T$ exhibits a rather irregular angular variation because these are in the crossover region between the type-I and  the type-II oscillations.

Figure 4  shows an enlarged plot for the angular dependence of $C(H, \phi)/T$
in the normal state at $T=0.65$~K in a field of 4~T.
Interestingly,
 the type-I oscillation 
 is also observed in the specific heat of the normal state.
 This fact implies that the magnetic-field response of the heavy quasiparticles in UBe$_{13}$ is anisotropic.
Very interestingly, a similar type-I angular variation of $C(H, \phi)/T$ is observed deep in the SC state at 0.08~K, suggesting a common origin for the anisotropy. 
 This type-I oscillation in the SC state is quite likely due to the quasiparticles in the vortex core, because of its similarity to the oscillation observed in the normal state. 
Having established that $C(H,\phi)/T$ in the SC state  exhibits rather unusual field and angular variation above 3~T, 
we consider it is not appropriate to discuss the nodal structure in this field region~\cite{Walti_PRBR_2001}.

The most striking outcome of the present study is the apparent absence of nodal quasiparticles in UBe$_{13}$.
The field and angular dependences of the specific heat in the SC state at low $T$ and  low $H$ strongly suggest that the gap is fully open over the whole Fermi surfaces, although the temperature variation of the specific heat is quite similar in appearance to that of a point-node gap.
A similar situation has been observed for the  heavy-electron superconductor CeCu$_2$Si$_2$~\cite{Kittaka_PRL_2014}, in which
the $T^2$-like variation of $C(T)$ at zero field is successfully reproduced by a two-band full-gap model.
It is, therefore, tempting to analyze the $C(T)$ of UBe$_{13}$ by a multi-band full-gap model,
employing the $\alpha$ model for 
 a strong-coupling superconductor~\cite{Padamsee_JLTP_1973}.
In order to reproduce $C(T)$, we need to assume three bands with different amplitudes of isotropic gaps. 
Figure 1(b) shows the fitting result with the parameters 
$(\gamma_1:\gamma_2:\gamma_3=55:38:7)$, and $(\alpha_1=1.9, \alpha_2=0.8, \alpha_3=0.3)$, where $\alpha_i=\Delta_i/(1.76k_{\rm B}T_{\rm c})$, and  $\Delta_{i}$  
 and $\gamma_i$ denotes  the SC gap and the electronic specific-heat coefficient of the $i$-th band, 
 respectively.
The simple model reproduces $C(T)$ of UBe$_{13}$ remarkably well.
Indeed, the calculated Fermi surfaces, consisting of two hole sheets and one electron sheet, support the occurrence of multi-band superconductivity in UBe$_{13}$~\cite{Takegahara_PhysicaB_2000, Maehira_PhysicaB_2002}.



\begin{centering}
\begin{figure}[t]
\includegraphics[width=8.4cm]{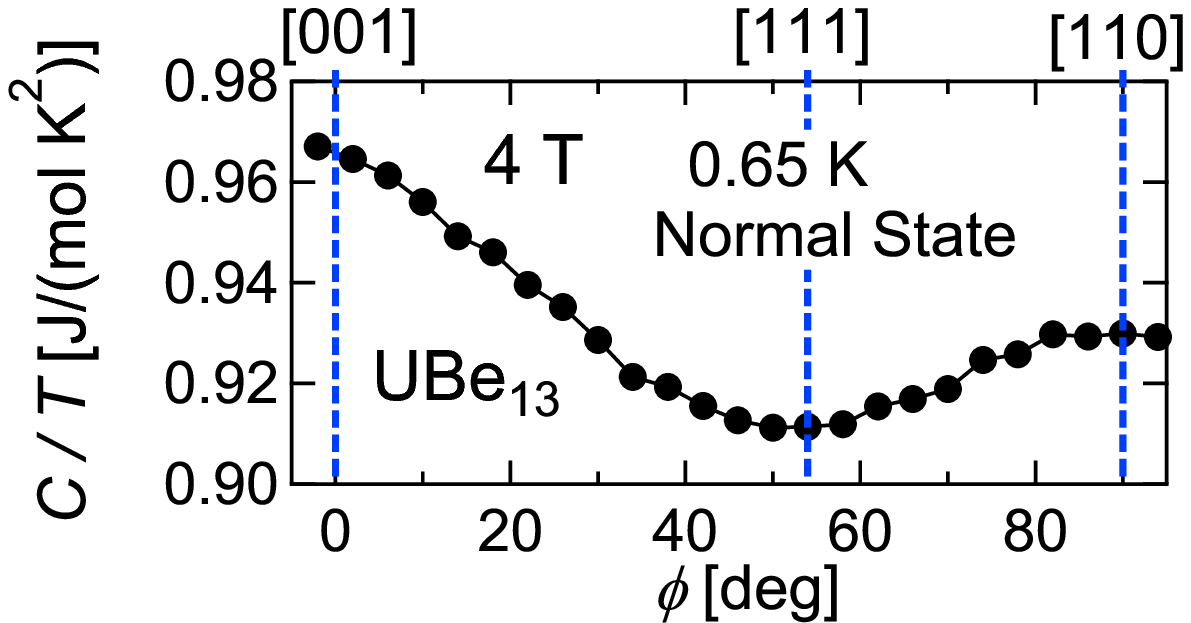}
\caption{   
Angular dependence of $C(H, \phi)/T$ in the normal state at $T=0.65$~K with $\mu_0H= 4$~T rotated in the $(1\bar{1}0)$ plane.
 }
\end{figure}
\end{centering}


There are two plausible scenarios for the absence of nodal quasiparticle excitations in UBe$_{13}$:
 (i) the SC gap function itself is nodeless, or
 (ii) the SC gap function has nodes only in the directions in which the Fermi surfaces are missing. 
The case (i) includes fully symmetric $A_{1g}$ representation. However,  for a cubic point group there are several possibilities for both even and odd parity unconventional pairing states in which the gap functions are nodeless~\cite{Volovik_JETP_1985, Blount_PRB_1985, Sigrist_RevModPhys_1991}. 
Regarding the case (ii), we would like to point out that, according to the band calculations, the Fermi surface is missing along the $\langle 111\rangle$ directions, except for a tiny electron band~\cite{Takegahara_PhysicaB_2000, Maehira_PhysicaB_2002}.
Hence,  
the quasiparticle excitation can be full-gap-like even if  \textit{point nodes exist} for the $\langle111\rangle$ directions.
Indeed, there are several odd-parity gap functions having only \textit{point nodes} along the 
$\langle111\rangle$ directions 
 \cite{Volovik_JETP_1985, Blount_PRB_1985, Sigrist_RevModPhys_1991,Ozaki_Machida_Ohmi_PTP_1985}.
 In either case, therefore, the absence of nodal quasiparticles does not necessarily rule out the possibility of unconventional pairing in UBe$_{13}$.

It is remarkable that $C(H,\phi)/T$ in the normal state exhibits a characteristic angular oscillation (Fig.~4).
In a cubic system, the Landau expansion of the free energy contains cubic invariants composed of magnetic-field components.
Accordingly, the field and angular dependences of the electronic specific heat can be expressed in the form 
$ C(T,H) \simeq C_{0}(T) + \alpha_{1}(T) H^2 + \alpha_{2}(T) H^4 
  + \beta(T) (H^{4}_{x} + H^{4}_{y} + H^{4}_{z})$, 
 where the last term represents the quartic invariant.
Note that the quadratic term is fully isotropic in cubic symmetry.
Therefore, the anisotropy emerges at a relatively high field, and $\beta > 0$  ($\beta < 0$) causes the type-I (type-II) oscillation.
In particular,  
 $ \frac{C_{ [001] } - C_{ [110] } }{ C_{ [110] } - C_{ [111]  } } 
 \simeq 3 $ should always hold, and 
is indeed almost satisfied   for  both the normal (Fig.~4) as well as the SC
 states  with the  type-I oscillation.
We also obtain  $\delta (C/T)$ = $(C_{  [001] }$ $-$ $C_{ [111] })/T$
 $\simeq \frac{2}{3} \beta H^4$, which  is roughly consistent with   
 the $\delta (C/T)$ at 0.95 K [Fig. 2(b)].

It is most plausible that the anisotropic response of $C(H)/T$ in the normal state comes from a Zeeman effect in heavy-electron bands.
The effective Fermi energy of UBe$_{13}$, $E_F/k_{\rm B}$, is reported to be $\sim$8~K~\cite{Ott_PRL_1984}. Therefore, a magnetic field of several tesla may result in a substantial Zeeman splitting of the heaviest band, leading to a reduction of the density of states at $E_F$. 
The data in Figs.~2(a) and 4 imply that this effect is anisotropic.
Unlike the case of Ce-based ($f^1$) compounds in which a magnetic Kondo effect plays a key role, there is no widely accepted model for the formation of a heavy-electron state of U-based ($f^2$) compounds, in particular of UBe$_{13}$ in which $C/T$ exhibits non-Fermi-liquid behavior down to near $T_{\rm c}$. 
In this regard, an intriguing candidate is a two-channel Kondo lattice model~\cite{Emery_PRB_1992,Anders_EurPhysJB_2002,Hoshino_PRL_2014}, in which odd frequency superconductivity has been predicted.
In any case, the observed anisotropic field response (Fig.~4) would provide a clue to unveil the unusual normal state in UBe$_{13}$.

Using $E_F/k_{\rm B}\sim$~8~K and  $\Delta/k_{\rm B}T_{\rm c}\sim$~3 for the heaviest band, we estimate the lowest excitation level for the localized quasiparticle state in the vortex core to be $\Delta^2/k_{\rm B}E_{\rm F}\sim$~1~K.
Thus, the quantum limit, in which a thermal broadening is narrower than the discrete bound state level 
 \cite{Caroli_PhysLett_1964, Hayashi_PRL_1998},  
may be realized at the base temperature (0.08~K) of the present experiment. 
In this situation, the core quasiparticles are nearly empty at least for the heaviest band, and would not contribute to a $H$-linear increase of $C(H)/T$. 
This effect can explain the unusual smallness of the initial slope of $C(H)/T$ in Fig.~1(c), with the $\gamma$ value of the heaviest band being roughly 1/2 of the total $\gamma$ as given by the multi-band analysis of $C(T)$ in Fig.~1(b).
We consider that the core quasiparticles are gradually recovered above $\sim$~3~T due to an overlapping of the discrete levels between neighboring vortices, and contribute to the observed anisotropy in $C(H,\phi)/T$ at 0.08~K.

Finally, we briefly discuss the possible origin of the $B^*$ anomaly.
In Ref.~\onlinecite{Kromer_PRL_1998}, $B^*$ has been discussed in the context of a precursor of a second phase transition in U$_{1-x}$Th$_x$Be$_{13}$. 
Here we propose an alternative scenario based on a Pauli paramagnetic effect.
The fact that $H_{\rm c2}(0)$ of UBe$_{13}$ is only $\sim 1/3$ of $H_{\rm c2}^{\rm orb}(0)$~\cite{Maple_PRL_1985} 
indicates that a significant paramagnetic suppression exists.
In a multi-band superconductor, the paramagnetic effect can be strongly band dependent, and a minor gap may be suppressed at a field $H^*$ much below $H_{\rm c2}(0)$. 
In this situation, a small peak would appear in $C(H)$ at $H^*$~\cite{Tsutsumi}, as observed in CeCu$_2$Si$_2$~\cite{Kittaka_PRL_2014}.
This $H^*$ might explain the $B^*$ anomaly.
It is also worthwhile to point out that the equilibrium magnetization of UBe$_{13}$ exhibits a minimum slightly below this field~\cite{YShimizu_PRL_2012}, possibly reflecting the Pauli paramagnetic effect~\cite{Tsutsumi}.

\color{black}
In conclusion, 
we studied  the SC symmetry and the low-energy quasiparticle excitations in  UBe$_{13}$ by means of
 specific-heat measurements on a high-quality single crystal.
The isotropic $H$-linear dependence of $C(H) $ in low fields and low temperatures   
indicates the absence of nodal quasiparticle excitations in the SC gap.
Whereas the present results do not exclude the possibility of unconventional pairing, the temperature dependence of the specific heat can be well explained by a multi-band full-gap model. 
A characteristic cubic anisotropy of $C(H)$ is observed at high fields with $H$ rotated in the  $(1\bar{1}0)$ plane, not only in the SC state but also in the normal state. 
This  might be a clue to understand the origin of NFL behavior as well as the enigmatic SC state in UBe$_{13}$.

We greatly appreciate valuable discussions with 
S. Hoshino, K. Miyake,  Y. Ikeda,  K. Izawa,  and 
  Y. Kuramoto.
The present work was supported in part by JSPS KAKENHI Grant Number 25103716, 26400360, 25800186.

\bibliography{apssamp}

\begin{thebibliography}{99}
\bibitem{Steglich_PRL_1979} F. Steglich, J. Aarts, C. D. Bredl, W. Lieke, D. Meschede, W. Franz, and H. Sch\"{a}fer, Phys. Rev. Lett. {\bf43}, 1892 (1979).
\bibitem{Ott_PRL_1983} H. R. Ott, H. Rudigier, Z. Fisk, and J. L. Smith, Phys. Rev. Lett. {\bf50}, 1595 (1983).
\bibitem{Kittaka_PRL_2014} S. Kittaka, Y. Aoki, Y. Shimura, T. Sakakibara, S. Seiro, C. Geibel, F. Steglich, H. Ikeda, and K. Machida, 
 Phys. Rev. Lett. {\bf112}, 067002 (2014).

\bibitem{Tien_PRB_1989} C. Tien, and I. M. Jiang, Phys. Rev. B {\bf40}, 229 (1989).

\bibitem{Tou07}H. Tou, N. Tsugawa, M. Sera, Y. Haga, and Y. Onuki, J. Magn. Magn. Mater. {\bf 310}, 706 (2007).
\bibitem{Sonier_PhysicaB_2003} J. E. Sonier, R. H. Heffner, G. D. Morris, D. E. MacLaughlin, O. O. Bernal, J. Cooley, J. L. Smith, J. D. Thompson, Physica B {\bf326}, 414 (2003).

\bibitem{Ott_PRL_1984} H. R. Ott, H. Rudigier, T. M. Rice, K. Ueda, Z. Fisk, J. L. Smith, 
 Phys. Rev. Lett. {\bf 52}, 1915 (1984).
\bibitem{Einzel_PRL_1986}
 D. Einzel,  P. J. Hirschfeld, F. Gross, B. S. Chandrasekhar, K. Andres, H. R. Ott, J. Beuers, Z. Fisk, and J. L. Smith,
 Phys. Rev. Lett. {\bf56}, 2513 (1986).
\bibitem{MacLaughlin_PRL_1984}
 D. E. MacLaughlin, C. Tien, W. G. Clark, M. D. Lan, Z. Fisk, J. L. Smith, H. R. Ott,
 Phys. Rev. Lett. {\bf 53}, 1833 (1984).
\bibitem{Golding_PRL_1985}
 B. Golding,  D. J. Bishop, B. Batlogg, W. H. Haemmerle, Z. Fisk, J. L. Smith, H. R. Ott,
 Phys. Rev. Lett. {\bf 55}, 2479 (1985).
\bibitem{Ellman_PRBR_1991} B. Ellman, T. F. Rosenbaum, J. S. Kim, and G. R. Stewart, Phys. Rev. B {\bf44}, 12074(R) (1991). 
\bibitem{Kromer_PRL_1998} F. Kromer, R. Helfrich, M. Lang, F. Steglich, C. Langhammer, A. Bach, T. Michels, J. S. Kim, and G. R. Stewart, Phys. Rev. Lett. {\bf81}, 4476 (1998). 
\bibitem{Walti_PRBR_2001} Ch.  W\"{a}lti, E. Felder, H. R. Ott, Z. Fisk, and J. L. Smith, Phys. Rev. B {\bf63},  R 100505 (2001).

\bibitem{YShimizu_PRL_2012} Y. Shimizu, Y. Haga, Y. Ikeda, T. Yanagisawa, and H. Amitsuka, Phys. Rev. Lett. {\bf109}, 217001 (2012).

\bibitem{Ott_PRBR_1985} H. R. Ott, H. Rudigier, Z. Fisk, and J. L. Smith, Phys. Rev. B {\bf31}, R 1651 (1985).;
 H. R. Ott,  H. Rudigier, E. Felder, Z. Fisk, and J. L. Smith, Phys. Rev. B {\bf33}, 126 (1986).

\bibitem{Mayer_PRB_1986} H. M. Mayer, U. Rauchschwalbe, C. D. Bredl, F. Steglich, H. Rietschel, H. Schmidt,
 H. W\"{u}hl, and J. Beuers, Phys. Rev. B {\bf33}, 3168 (1986).
\bibitem{Gegenwart_PhysicaC_2004} P. Gegenwart, C. Langhammer, R. Helfrich, N. Oeschler, M. Lang, 
 J. S. Kim, G. R. Stewart, and F. Steglich, Physica C {\bf408}-{\bf410}, 157 (2004).

\bibitem{McElfresh_PRB_1993} M. McElfresh, M. B. Maple, J. O. Willis, D. Schiferl, J. L. Smith, Z. Fisk, and D. L. Cox, Phys. Rev. B {\bf48}, 10395 (1993).

\bibitem{Cox_PRL_1987} D. L. Cox, Phys. Rev. Lett. {\bf59}, 1240 (1987).

\bibitem{Schmiedeshoff_PhilosMagazine_2009}
 G. M. Schmiedeshoff, and J. L. Smith, Philos. Mag. {\bf89}, 1839 (2009).
\bibitem{Nishiyama_JPSJ_2011} S. Nishiyama, and K. Miyake, J. Phys. Soc. Jpn. {\bf79}, 104711 (2011). 

\bibitem{Haga_Physica_1999} Y. Haga, E. Yamamoto, T. Honma, A. Nakamura, M. Hedo, S. Araki, H. Ohkuni, and Y. Onuki,
 Physica  {\bf259B}, 627 (1999). 

\bibitem{Volovik_JETP_Lett_1993}  G. E. Volovik, JETP. Lett. {\bf58}, 469 (1993).
\bibitem{Vekhter_PRBR_1999} I. Vekhter, P. J. Hirschfeld, J. P. Carbotte, and E. J. Nicol, Phys. Rev. B {\bf59}, R 9023 (1999).
\bibitem{Sakakibara_JPSJ_2007} T. Sakakibara, A. Yamada, J. Custers, K. Yano, T. Tayama, H. Aoki, and K. Machida,
 J. Phys. Soc. Jpn. {\bf76}, 051004 (2007).  

\bibitem{Volovik_JPhysC_21_1988} G. E. Volivik, J. Phys. C: {\bf21}, L221 (1988).
\bibitem{Miranovic_PRB_2003} P. Miranovi\'{c}, N. Nakai, M. Ichioka, and K. Machida, Phys. Rev. B {\bf68}, 052501 (2003).

\bibitem{Nakai_PRBR_2004} N. Nakai, P. Miranovi\'{c}, M. Ichioka, and K. Machida, Phys. Rev. B {\bf70}, R 100503 (2004).

\bibitem{Volovik97} G.E. Volovik, JETP Lett. {\bf 65}, 491 (1997).

\bibitem{Maple_PRL_1985} M. B. Maple,  J. W. Chen, S. E. Lambert, Z. Fisk, J. L. Smith, H. R. Ott, J. S. Brooks, and M. J. Naughton, 
  Phys. Rev. Lett. {\bf54}, 477 (1985).
\bibitem{YShimizu_DrThesis_2012} Y. Shimizu, Ph.D. thesis, Hokkaido University, Sapporo, (2012).
\bibitem{YShimizu_JPSconfProc_2014} Y. Shimizu, Y. Haga, T. Yanagisawa, H. Amitsuka, D. Aoki, J-P. Brison, and D. Braithwaite, J. Phys. Soc. Conf. Proc. {\bf3}, 015009 (2014).
\bibitem{Padamsee_JLTP_1973} H. Padamsee, J. E. Neighbor, and C. A. Shiffman, J. Low. Temp. Phys. {\bf12} (1973) 387.
\bibitem{Takegahara_PhysicaB_2000} K. Takegahara, and H. Harima, Physica B {\bf281}, 764 (2000).
\bibitem{Maehira_PhysicaB_2002} T. Maehira, A. Higashiya, M. Higuchi, H. Yasuhara, and A. Hasegawa, Physica B {\bf312-313}, 103 (2002).

\bibitem{Volovik_JETP_1985} G. E. Volovik, and L. P. Gor'kov, JETP {\bf61}, 843 (1985).
\bibitem{Blount_PRB_1985} E. I. Blount, Phys. Rev. B {\bf32}, 2935 (1985).
\bibitem{Sigrist_RevModPhys_1991} M. Sigrist, and K. Ueda, Rev. Mod. Phys. {\bf63}, 239 (1991). 

\bibitem{Ozaki_Machida_Ohmi_PTP_1985} M. Ozaki, K. Machida, and T. Ohmi, Prog. Theor. Phys. {\bf74}, 221 (1985).

\bibitem{Emery_PRB_1992} V. J. Emery, and S. Kivelson, Phys. Rev. B {\bf46}, 10812 (1992).
\bibitem{Anders_EurPhysJB_2002} F. B. Anders, Eur. Phys. J. B {\bf28}, 9 (2002).
\bibitem{Hoshino_PRL_2014}  S. Hoshino, and Y. Kuramoto, Phys. Rev. Lett. {\bf112}, 167204 (2014).
\bibitem{Caroli_PhysLett_1964} C. Caroli, P. G. de Gennes, and J. Matricon, Phys. Lett. {\bf9}, 307 (1964).
\bibitem{Hayashi_PRL_1998} N. Hayashi, T. Isoshima, M. Ichioka, and K. Machida, Phys. Rev. Lett. {\bf80}, (1998) 2921.
\bibitem{Tsutsumi}Y. Tsutsumi, K. Machida, and M. Ichioka, arXiv:1410.3555.

\end{thebibliography}


\end{document}